\documentclass[twocolumn,aps,prd,amsmath,amssymb,nofootinbib]{revtex4-1}
\usepackage{graphicx,bm,tikz,amsmath}
\newcommand{\beq}{\begin{equation}}
\newcommand{\bea}{\begin{eqnarray}}
\newcommand{\eeq}{\end{equation}}
\newcommand{\eea}{\end{eqnarray}}
\newcommand{\bal}{\begin{align}}
\newcommand{\eal}{\end{align}}
\newcommand{\mx}{m_\chi}

\newcommand{\be}{\begin{equation}}
\newcommand{\ee}{\end{equation}}

\begin{document}

\title{Galactic Center Excess in Gamma Rays from Annihilation of Self-Interacting Dark Matter}

\author{Manoj Kaplinghat}
\affiliation{Department of Physics and Astronomy, University of California, Irvine, CA 92697, USA}

\author{Tim Linden}
\affiliation{Kavli Institute for Cosmological Physics, University of Chicago, IL 60637, USA}

\author{Hai-Bo Yu}
\affiliation{Department of Physics and Astronomy, University of California, Riverside, CA 92521, USA}

\begin{abstract}
Observations by the Fermi-LAT telescope have uncovered a significant $\gamma$-ray excess toward the Milky Way Galactic Center. There has been no detection of a similar signal in the direction of the Milky Way dwarf spheroidal galaxies. Additionally, astronomical observations indicate that dwarf galaxies and other faint galaxies are less dense than predicted by the simplest cold dark matter models. We show that a self-interacting dark matter model with a particle mass of roughly 50 GeV annihilating to the mediator responsible for the strong self-interaction can simultaneously explain all three observations. The mediator is necessarily unstable and its mass must be below about 100 MeV in order to lower densities in faint galaxies. If the mediator decays to electron-positron pairs with a cross section on the order of the thermal relic value, then we find that these pairs can up-scatter the interstellar radiation field and produce the observed $\gamma$-ray excess. We show that this model is compatible with all current constraints and highlight detectable signatures unique to self-interacting dark matter models.
\end{abstract}

\maketitle

\noindent {\bf The Galactic Center excess}. Recent Fermi-LAT observations of the Galactic Center (GC) of the Milky Way have uncovered a stunning $\gamma$-ray excess compared to expectations from diffuse astrophysical emission~\cite{Goodenough:2009gk, Hooper:2010mq, Hooper:2011ti, Abazajian:2012pn, Gordon:2013vta, Macias:2013vya, Abazajian:2014fta, Daylan:2014rsa, Zhou:2014lva, Calore:2014xka, Abazajian:2014hsa}. While these studies differ in the astrophysical background models, they all agree on three key features of the $\gamma$-ray excess:
(1) the spectrum is strongly peaked at an energy of approximately 2~GeV, with a low-energy spectrum that is harder than expected from $\pi^0$-emission, (2) the excess radially extends to at least 10$^\circ$ from the GC, following an emission profile that falls with distance ($r$) from the GC as $r^{-\alpha}$ with $\alpha$~=~2.0~--~2.8, and (3) the excess is roughly spherically symmetric, without any evidence of elongation parallel or perpendicular to the galactic plane. 

While other explanations have been discussed~\cite{Abazajian:2010zy, Abazajian:2012pn, Gordon:2013vta, Petrovic:2014xra,Carlson:2014cwa,Petrovic:2014uda}, dark matter remains a compelling possibility. The detection of an excess with the same spectrum toward dwarf spheroidal galaxies surrounding the Milky Way would verify this possibility. However, no equivalent signal has been observed in dwarf spheroidal galaxies~\citep{Geringer-Sameth:2014qqa}, which stands in mild tension with some models of the GC excess~\citep{fermi_symposium_dwarfs}.

In fact, dwarf galaxies have long challenged our understanding of the nature of dark matter. The dark matter halos of dwarf galaxies have constant density cores~\cite{Moore:1994yx,Flores:1994gz, Walker:2011zu, KuziodeNaray:2007qi}, in contrast to the cuspy profile predicted by simulations of cold collisionless dark matter (CDM). Additionally, CDM predicts a population of dwarf halos that are systematically denser than the dwarf spheroidal galaxies in the Milky Way~\cite{BoylanKolchin:2011dk}, Andromeda~\cite{Tollerud:2014zha}, or Local Group~\cite{Kirby:2014sya,Garrison-Kimmel:2014vqa}. A compelling solution to these challenges is to assume that dark matter strongly interacts with itself~\cite{Spergel:1999mh,Firmani:2000ce}. Recent simulations have shown that nuclear-scale dark matter self-interaction cross sections can produce heat transfer from the hot outer region to the cold inner region of dark matter halos, reducing the central densities of dwarf galaxies in accordance with observations~\cite{Vogelsberger:2012ku,Rocha:2012jg,Zavala:2012us,Elbert:2014bma}. 

\noindent {\bf Connection to dark matter self-interactions.} We explore the intriguing possibility that the GC $\gamma-$ray excess is caused by Inverse Compton (IC) scattering of energetic e$^+$e$^-$ from dark matter annihilation, and the absence of the GeV $\gamma$-ray signal in dwarf spheroids is a natural consequence of self-interacting dark matter (SIDM) models. Our key observations are as follows:

\begin{itemize}

\item{Energetic e$^+$e$^-$ from dark matter annihilation (or another extended source distribution) can effectively produce $\gamma$-rays in the GC through IC and bremsstrahlung, due to the high interstellar radiation field (ISRF) and gas densities in this region. The IC emission can explain the peak of the GC signal (at 2-3 GeV) for dark matter masses in the approximate mass range of 20-60 GeV. The crucial requirement is the presence of a new source of e$^+$e$^-$ with energies larger than $20$ GeV, which produce $\gamma$-rays with peak energy of $\sim (20\ {\rm GeV}/m_e)^2 E_{\rm ISRF}$, with typical ISRF photon energy $E_{\rm ISRF} \sim 1\ {\rm eV}$.}

\item{The AMS-02 constraint \cite{Bergstrom:2013jra} demands a softer electron spectrum than direct annihilation to e$^+$e$^-$ and hence annihilation through a light mediator is a natural solution\footnote{Another possibility, which we do not explore here, is direct annihilation to  $\mu^+\mu^-$, with $e$ and $\tau$ channels suppressed.}.}

\item{A nuclear-scale dark matter self-scattering cross section requires a dark force carrier with a mass below $\sim100$ MeV~\cite{Tulin:2013teo,Cline:2013zca,Boddy:2014yra}. Annihilations through this mediator can kinematically couple only to e$^+$e$^-$ and neutrinos in the standard model sector.
}

\item{The e$^+$e$^-$ produced via dark matter annihilation do not produce appreciable $\gamma$-rays from dwarf galaxies due to their low starlight and gas densities. The dominant signal in these systems should be due to final state radiation, with a cross section suppressed by $\alpha_{\rm EM}$.}

\item{The SIDM density profile in the central region of the Milky Way is determined by the bulge potential~\citep{Kaplinghat:2013xca}. Models of the galactic bulge imply that the dark matter density increases to within 1-2$^\circ$ from the GC and the annihilation power is significantly enhanced compared to the predictions of SIDM-only simulations.}

\end{itemize}

\noindent {\bf A hidden sector dark matter model}. 
We consider a simple hidden sector model in which a 50~GeV dark matter particle couples to a vector mediator $\phi$. The relic density in this model is set by the annihilation process $\bar{\chi}\chi\rightarrow\phi\phi$ with an annihilation cross section $4.4 \xi \times10^{-26}~{\rm cm^3 s^{-1}}$ for a Dirac particle~\cite{Steigman:2012nb}, where $\xi$ is the ratio of the temperature of the hidden sector to that of the visible sector at freeze-out \cite{Feng:2009mn}. We assume that the hidden and visible sectors are coupled through kinetic~\cite{Holdom:1985ag} or Z-mixing~\cite{Babu:1996vt} leading to $\phi\rightarrow$ e$^+$e$^-$ decays~\cite{Kaplinghat:2013yxa}. The dark matter mass utilized here illustrates our main points; the mass could be large or smaller by about 50\%, depending on the details of electron energy loss in the GC. 

In order to compute the secondary emission from this model we utilize the software {\tt PPPC4DMID} \citep{Cirelli:2010xx}, which provides the solution for one-dimensional diffusion with spatially dependent energy losses. We use the ``MED" diffusion parameters listed in {\tt PPPC4DMID} \citep{Cirelli:2010xx}. This software calculates the $\gamma$-ray spectrum from IC scattering assuming an interstellar radiation field energy density from {\tt GALPROP} \cite{Vladimirov:2010aq}, an exponential magnetic field profile \cite{Strong:1998fr}, and negligible bremsstrahlung losses, which is a good approximation for the $\gtrsim 10$ GeV electrons under consideration. We tested the {\tt PPPC4DMID} spectrum by writing an independent code that solves the one-dimensional diffusion equation assuming spatially-constant energy losses and found good agreement.

\begin{figure}
\begin{tabular}{c}
\includegraphics[scale=.32]{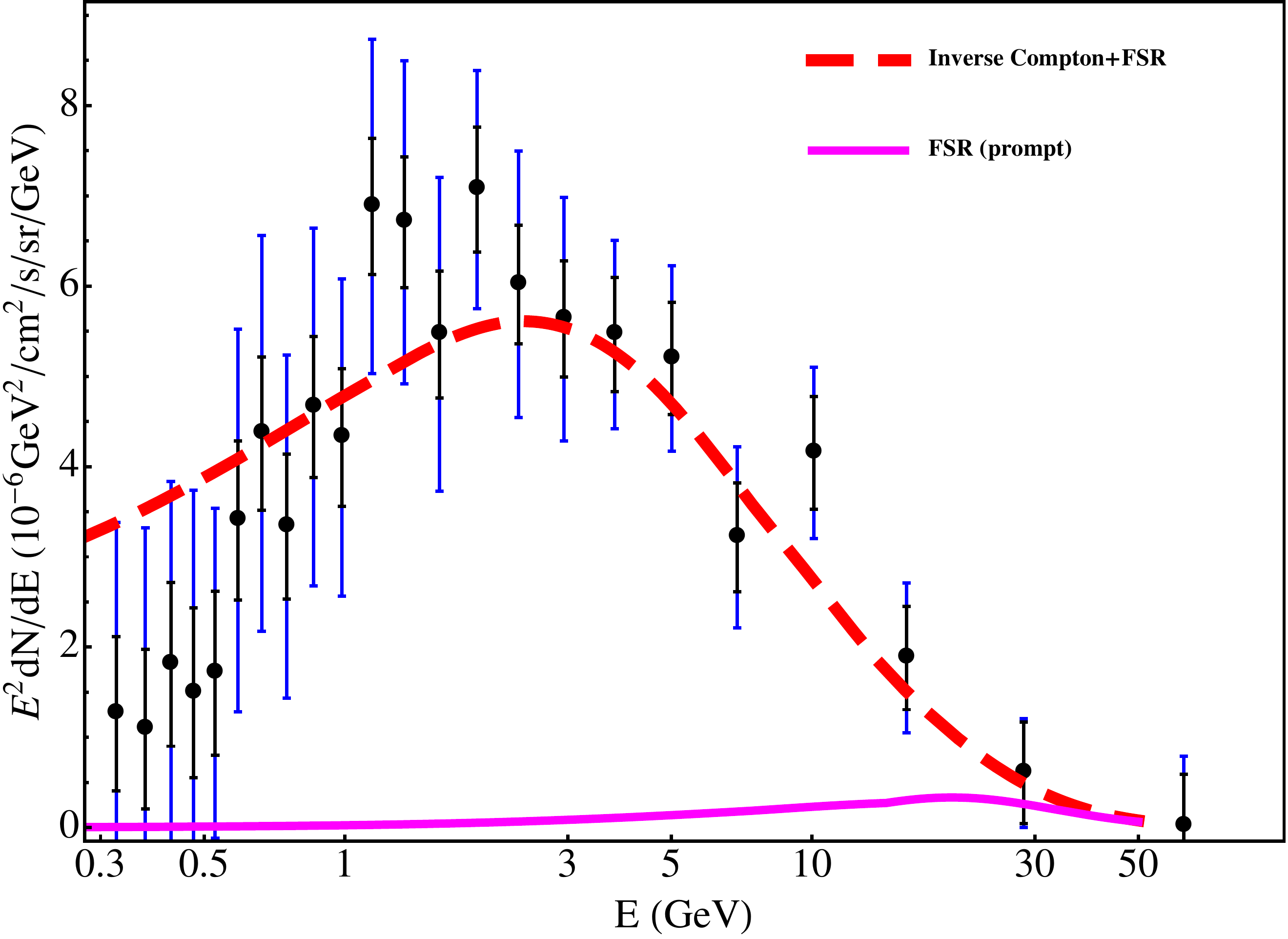}
\end{tabular}
\caption{$\gamma$-ray spectrum from Inverse Compton emission and final state radiation produced by annihilation of a 50~GeV dark matter particle through a light mediator into e$^+$e$^-$ final state. The spectrum is compared to the Galactic Center excess~\cite{Calore:2014xka}.}
\label{fig:gc_spectrum}
\end{figure}

\begin{figure}
\begin{tabular}{c}
\includegraphics[scale=1.0]{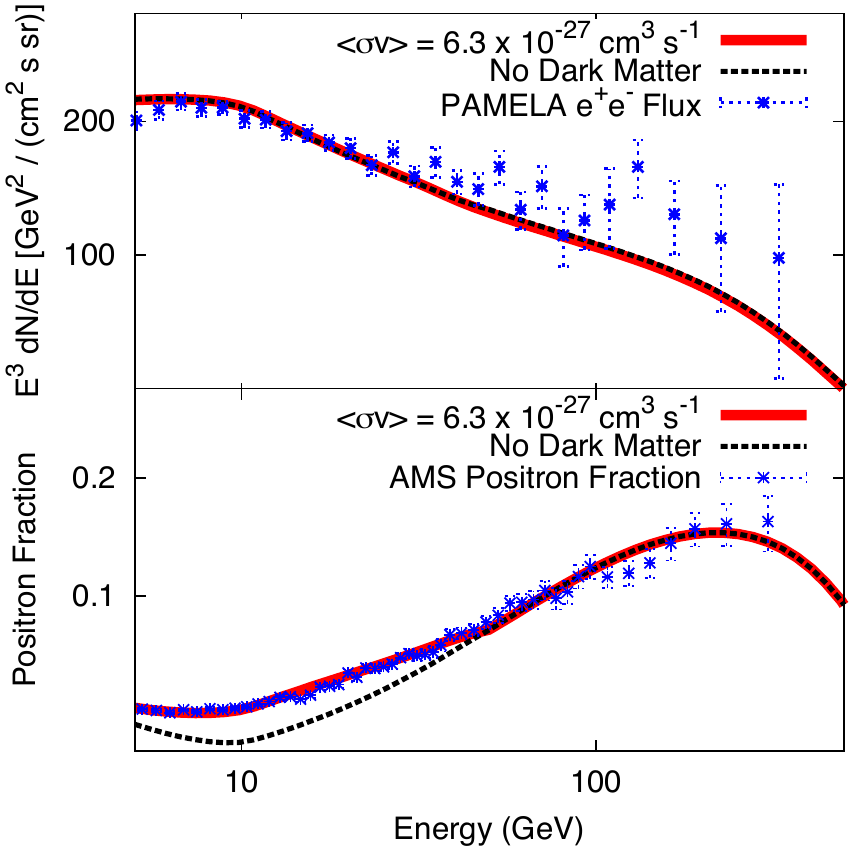}
\end{tabular}
\caption{The e$^+$e$^-$ spectrum (top) and positron fraction (bottom) for the SIDM model, compared to observations from PAMELA and AMS-02, respectively.}
\label{fig:ams_positrons}
\end{figure}

In Fig.~\ref{fig:ams_positrons}, we compare the intensity and spectrum from our model to the region of interest (ROI) 1 of Ref.~\cite{Calore:2014xka}, which includes regions within $5^\circ$ radius (about 750 pc projected distance at the GC) excluding latitudes $|b|~<~2^\circ$, finding good agreement. 

These results show that the secondary IC emission effectively reproduces the hard spectral bump at an energy of $\sim$2~GeV, and the relatively hard spectrum component at energies above 10~GeV observed by Ref.~\cite{Calore:2014xka}. The hard spectrum component is an important discriminator of the dark matter mass, as it absent for masses closer to 20 GeV. 

\begin{figure}
\begin{tabular}{c}
\includegraphics[scale=.33]{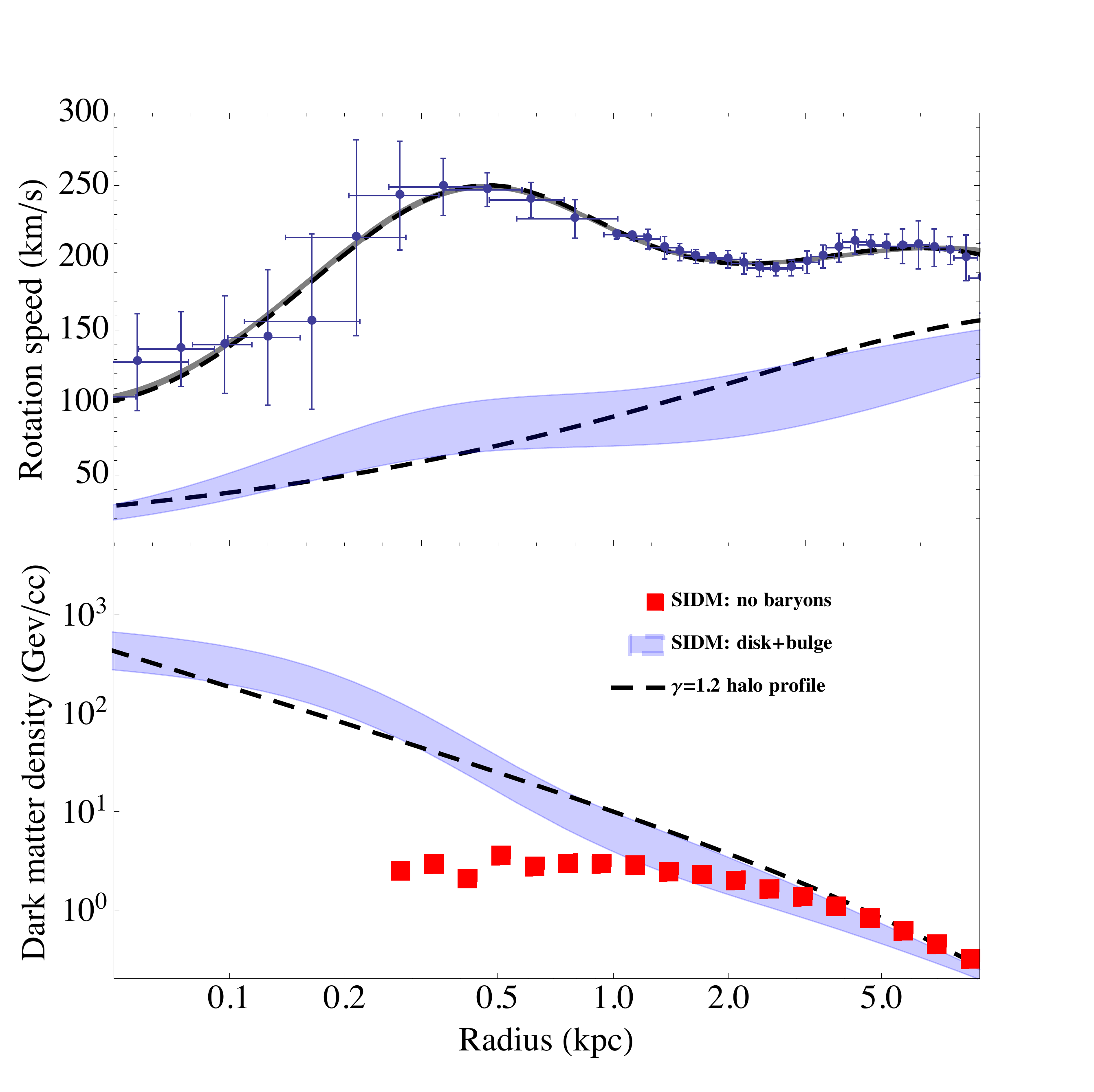}
\end{tabular}
\caption{The top panel shows the rotation curve data for the Milky Way complied by Ref. \cite{Sofue:2013kja} and fits described in the text. In the panel below, the adopted $\gamma=1.2$ density profile (dashed) is compared to SIDM predictions (shaded), with the filled points showing the density profile predicted for SIDM without including baryons \cite{Rocha:2012jg}.}
\label{fig:rot_curve}
\end{figure}

We estimate the range of cross sections required to produce this signal as $0.3-2 \times 10^{-26} {\rm cm}^3/{\rm s}$, corresponding to the SIDM density profiles shown in Fig.~\ref{fig:rot_curve} and discussed next. In order to estimate this cross section range we noted (using the density profiles available in {\tt PPPC4DMID}) that the IC signal (shown in Fig.~\ref{fig:ams_positrons}) is proportional to the $J$-factor ($J=\int d\ell \rho^2(\ell,\Omega)$, where $\ell=$ line of sight) within 5 degrees of the GC at the  10\% accuracy level. Therefore, we scale the {\tt PPPC4DMID} result using the $J$-factors for the SIDM density profiles to obtain the cross section range.

{\noindent {\bf Density profile of SIDM}.}
The cross section estimate depends on the dark matter density profile. The expectation from SIDM-only simulations is that the SIDM density profile would be essentially constant in this region. However, when baryons dominate, as expected in the inner galaxy, it has been shown that the equilibrium SIDM density profile tracks the baryonic potential~\cite{Kaplinghat:2013xca}. We compute the equilibrium SIDM density profile assuming two possibilities for the early (before self-interactions become effective) dark matter density profile following the method in Ref.~\cite{Kaplinghat:2013xca}: an NFW profile \cite{Navarro:1996gj} with scale factor $r_s=26$ kpc \cite{Prada:2011jf} and the same profile after adiabatic contraction \cite{Blumenthal:1985qy} due to the disk and bulge of the Milky Way. The early profiles set the boundary conditions for the equilibrium solutions.

We determine the consistency of this scenario by fitting to the composite galactic rotation curve of Ref.~\cite{Sofue:2013kja} shown in the top panel of Fig.~\ref{fig:rot_curve} including a black hole of mass $4\times 10^6 {\rm M}_\odot$, inner and outer spherical bulges with exponential density profiles, an exponential disk and a spherical halo. The SIDM halo is computed self-consistently from the spherically-averaged stellar distribution. The general non-spherical equilibrium solution  \cite{Kaplinghat:2013xca,Amorisco:2010sb} would require non-spherical modeling of the bulge, which is beyond the scope of the present work. 

A simple iterative procedure suffices to find the best fit values after varying the outer bulge and disk parameters. The inner bulge and black hole mass are not varied as the SIDM halo profile outside 100 pc is less sensitive to them. The SIDM halo has two free parameters: central density ($\rho_0$) and central velocity dispersion ($\sigma_0$). We fix $\sigma_0=170$ km/s and vary $\rho_0$ in the interative procedure described above. Equilibrium solutions also exist for slightly different values of $\sigma_0$, but they don't change our key results. 

The SIDM fits and the density profiles are shown in Fig.~\ref{fig:rot_curve}, with the edges of the shaded band arising from the two assumptions about the early dark matter profile. For comparison, we also show the fit to the composite rotation curve with an NFW-like dark matter profile $\rho \propto r^{-\gamma}(1+r/r_s)^{\gamma-3}$ with $\gamma=1.2$, $r_s=10$ kpc and local density of 0.3 ${\rm GeV}/{\rm cm}^3$. We choose this profile since $\gamma=1.2$ is consistent with the GC excess fits  \cite{Macias:2013vya,Abazajian:2014fta,Daylan:2014rsa,Calore:2014xka} and it also closely tracks the adibatically contracted profile mentioned previously. 

Our inner bulge parameters are consistent with Ref.~\cite{Sofue:2013kja}. The main bulge has an exponential scale radius of approximately $0.13~{\rm kpc}$ and total mass of $8-9 \times 10^9 {\rm M}_\odot$ depending on the model, again consistent with Ref.~\cite{Sofue:2013kja}. The scale radius is significantly smaller than the bulge radius found in photometric studies of the bulge \cite{Cao:2013dwa,Wegg:2013upa,Saito:2011zq}, perhaps due to additional bulge structures in the inner 0.3 kpc, for which there is indirect evidence \cite{Gonzalez:2011kf}. For our purposes, the present model suffices to convert the observed rotation curve to a baryonic potential, which then determines the SIDM density profile. 

The main point of this exercise is to emphasize that SIDM predicts high dark matter densities at the Milky Way center (and in other baryon-dominated galaxies) and to explicitly show that these dark matter densities are compatible with current rotation curve data. Thus, the SIDM $J$-factor for annihilation is comparable to or larger than the CDM predictions. 

\noindent {\bf Morphology of the excess.} Unlike the case of the prompt signal, which is proportional to the annihilation power (density squared), the IC morphology is affected by diffusion and energy losses. In addition, different studies of the GC and inner galaxy \cite{Macias:2013vya,Abazajian:2014fta,Daylan:2014rsa,Calore:2014xka} have inferred somewhat different spatial templates for the signal. These facts suggest that signal morphology may be an important discriminant but more work is required.

We note that despite the interaction of the e$^+$e$^-$ flux with the non-spherical ISRF, the resultant IC signal (using {\tt PPPC4DMID}) is roughly spherically symmetric (to within 10\% in the ROI), in agreement with present estimates~\cite{Daylan:2014rsa}. Diffusion and energy losses may also moderate any asphericity in the annihilation power, suggesting that the sphericity of the GeV excess can be accommodated in this SIDM scenario. 

In comparison, for a prompt signal to be spherical, one must appeal to significant gas cooling in the halo centers after all the major mergers \cite{Kazantzidis:2004vu} since CDM halos without baryons are highly aspherical in their centers \cite{Allgood:2005eu}. However, adiabatic contraction due to this cooling may result in $\gamma \approx 2$ \cite{Zemp:2011nk}, which would be inconsistent with the GeV excess templates. There seems to be no clear prediction for the morphology of prompt gamma-rays from dark matter annihilation.

{\noindent {\bf AMS-02 constraint}.} The strongest constraint on SIDM models stems from observations of the local positron fraction by PAMELA~\citep{Adriani:2008zr} and AMS-02~\citep{Aguilar:2013qda}. While leptophilic dark matter was originally posited as a solution to the rising positron fraction~\cite{Cirelli:2008jk,ArkaniHamed:2008qn,Pospelov:2008jd}, these models have come into tension with constraints from the Fermi-LAT~\citep{Hooper:2012sr}, H.E.S.S.~\citep{Abazajian:2011ak} and PLANCK~\citep{Slatyer:2009yq}. Instead, working within the framework that pulsars produce the rising positron fraction~\citep{Hooper:2008kg,Profumo:2008ms,Linden:2013mqa}, AMS-02 observations can set strong limits on the annihilation of dark matter to leptophilic final states. For dark matter annihilation directly to e$^+$e$^-$ pairs, these limits can fall below the thermal cross section by two orders of magnitude~\citep{Bergstrom:2013jra}. However, we show below that these constraints are relaxed considerably if the e$^+$e$^-$ pairs are produced in the decay of a mediator because of the softer energy spectrum.

In Fig.~\ref{fig:ams_positrons}, we produce \emph{Galprop} models~\citep{Strong:1998pw} of the astrophysical electron and positron flux at Earth. We add a pulsar component modeled by a power-law injection of e$^+$e$^-$ pairs with an exponential cutoff~\citep{Profumo:2008ms}. We add a dark matter annihilation component assuming the hidden and visible sectors are coupled through Z-boson mass mixing, with $\xi=1$ and an annihilation cross section to e$^+$e$^-$ of $4.4 \xi \times10^{-26}~{\rm cm^3 s^{-1}}/7=6.3\times10^{-27}~{\rm cm^3 s^{-1}}$~\cite{Kaplinghat:2013yxa}.  This cross section is within the range computed previously for SIDM models that explain the GC GeV excess. We use the $\gamma=1.2$ profile with a local density of 0.3~GeV~cm$^{-3}$ shown in Fig.~\ref{fig:rot_curve} since it tracks the upper edge of the shaded band near the solar location and results in a conservative AMS-02 constraint. Because the best-fit \emph{Galprop} diffusion parameters calculated by primary-to-secondary observations~\citep{Trotta:2010mx} may not represent the average diffusion parameters for leptons, we test an ensemble of diffusion parameters and find the model producing the best combined fit to the AMS-02 positron fraction and the PAMELA e$^+$e$^-$ flux. The resulting diffusion parameters are not far from those calculated for cosmic-ray nuclei in \citep{Trotta:2010mx}. Specifically we use a diffusion constant of 9.1~$\times$~10$^{28}$~cm$^2$s$^{-1}$, a half-scale height of 6.6~kpc, an Alfv{\'e}n velocity of 30.5~km~s$^{-1}$, and a primary cosmic-ray electron spectrum following a broken power-law falling as E$^{-2.23}$ below 11.4~GeV and as E$^{-2.79}$ at higher energies. We adopt charge-dependent solar modulation, with amplitudes of $\phi_{e^{+}}$~=~171~MV for positrons and $\phi_{e^{-}}$~=~54~MV for electrons.

This result shows that annihilations through a light mediator can reproduce the intensity of the GC excess while remaining consistent with AMS-02 constraints. We note that the formal fit for AMS-02 data is quite good ($\chi^2$/d.o.f.~=~1.32), but poor for PAMELA data ($\chi^2$/d.o.f.~=~3.07). However, we find this to be of a similar quality to fits produced without a dark matter component. Interestingly, updated measurements by AMS-02 offer the exciting possibility of constraining or detecting SIDM annihilation. We note that these findings are consistent with \citep{Bergstrom:2013jra}, given that we adopt a local density of 0.3~GeV~cm$^{-3}$, as opposed to 0.4~GeV~cm$^{-3}$, and noting that the annihilations through a light mediator soften the resulting e$^+$e$^-$ injection spectrum, making it more comparable to annihilation through $\mu^+\mu^-$ than to direction annihilation to e$^+$e$^-$ pairs. 

\begin{figure}
\includegraphics[scale=1.0]{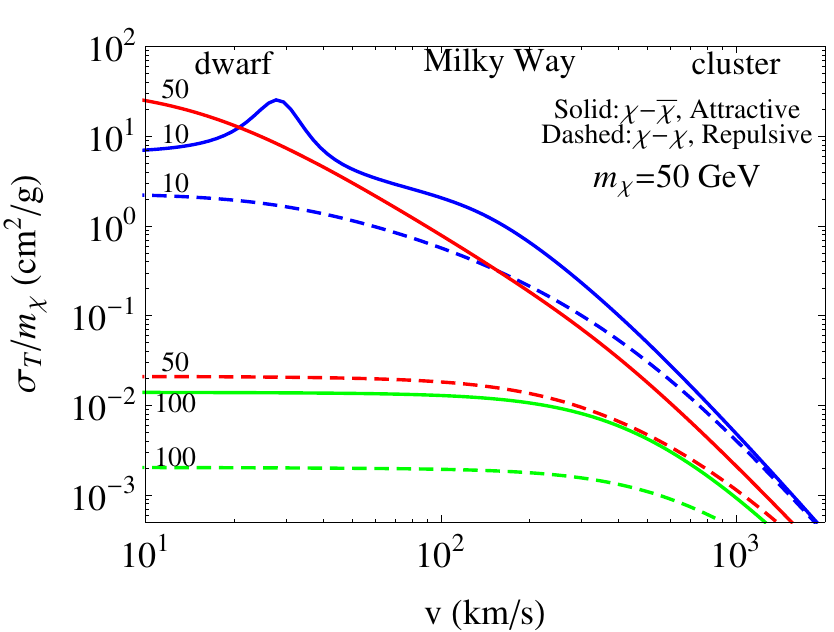}
\caption{Dark matter self-interaction cross section as a function of the dark matter relative velocity for mediator mass 10 MeV (Blue), 50 MeV (red), and 100 MeV (Green). SIDM models with $\sigma_T/\mx\sim0.5-50~{\rm cm^2/g}$ on dwarf scales can produce constant density cores in dwarf galaxies in accordance with observations~\cite{Elbert:2014bma}.}
\label{fig:sidm}
\end{figure}

\noindent {\bf SIDM solution to small-scale structure formation problems}.
In Fig.~\ref{fig:sidm}, we show the dark matter self-interaction cross section as a function of the dark matter relative velocity. For 50~GeV dark matter, the mediator must be less than 50 MeV for self-interactions to solve anomalies on dwarf scales. It is interesting to note that for attractive interactions $\sigma_T$ is enhanced when the mediator mass is $50$ MeV because of the s-wave resonance~\cite{Tulin:2012wi}. The cross section $\sigma_T$ drops slightly on Milky Way scales, but is still large enough to effect the Milky Way halo~\cite{Kaplinghat:2013xca}. On cluster scales, dark matter self-scattering is highly suppressed as $1/v^4$, because the momentum transfer is much larger than the mediator mass and dark matter self-scattering occurs in the Rutherford scattering limit. Therefore, the model is fully consistent with constraints $\sigma_T/\mx\lesssim1~{\rm cm^2/g}$ for $v\sim3000~{\rm km/s}$ from the Bullet Cluster~\cite{Randall:2007ph} and cluster shape constraints \cite{Peter:2012jh}.

\noindent {\bf Other detectable features}.
In addition to producing a correlation between the density profile of dark matter and the galactic bulge, as well as producing a significant contribution to the AMS-02 positron fraction, the dark matter model described here may be tested through radio observations. Specifically, the large e$^+$e$^-$ flux predicted by our model may be able to explain (or be constrained by) the Green Bank Telescope radio continuum observations towards the GC \cite{YusefZadeh:2012nh}, the isotropic emission detected by ARCADE-2~\cite{Fixsen:2009xn, Fornengo:2011cn, Hooper:2012jc, Fang:2014joa} and the observation of hard-spectrum radio filaments in the GC~\cite{Linden:2011au}.

Additionally, the proposed SIDM particle can scatter with nuclei through the mediator $\phi$, leading to direct detection signals~\cite{Kaplinghat:2013yxa}. Since the mediator mass is comparable to the momentum transfer of nuclear recoils, the event spectrum has a non-trivial momentum dependence~\cite{Kaplinghat:2013yxa, Fornengo:2011sz}, which could provide a smoking-gun signature for SIDM. 

\noindent {\bf Summary}. 
We show that the GC excess can be explained through secondary emission from e$^+$e$^-$ pairs produced in dark matter annihilation events, a scenario which naturally predicts suppressed $\gamma$-ray emission from dwarf spheroidal galaxies. This class of models is well-motivated in the context of SIDM models posited to explain anomalies in the dark matter density profiles of dwarf spheroidal galaxies and other baryon-poor galaxies. These models make unique predictions, which could be tested in the near future.

While this paper was in preparation, several related papers were submitted by other  groups~\citep{Calore:2014nla, Liu:2014cma}. We note that our favored SIDM mass range is consistent with the analysis of light-mediator models by~\citep{Calore:2014nla}. 

\emph{Acknowledgements -} We thank Francesca Calore for providing the data employed in Fig.~\ref{fig:gc_spectrum}. MK is supported by NSF Grant No. PHY-1214648. TL is supported by the National Aeronautics and Space Administration through Einstein Postdoctoral Fellowship Award Number PF3-140110. HBY is supported by startup funds from UCR. We thank the Aspen Center for Physics and the NSF Grant-1066293 for hospitality during the conception of this paper.

\bibliographystyle{h-physrev.bst}

\bibliography{sidm}

\end{document}